\documentclass[aps,twocolumn,amsmath,amssymb,superscriptaddress,prl,longbibliography]{revtex4-2}
\usepackage{graphicx,color,bm}
\usepackage[english]{babel}
\usepackage{amsmath,amssymb}
\usepackage{dsfont}
\usepackage{subfigure}
\usepackage{epstopdf}
\usepackage{soul}
\usepackage{xcolor}
\usepackage[normalem]{ulem}
\usepackage{cancel}
\usepackage[linktocpage=true,colorlinks=true,pdfborder={0 0 0},linkcolor=blue,citecolor=blue,filecolor=yellow,urlcolor=blue,bookmarks,pdfauthor={},]{hyperref}

\newcommand{\denis}[1]{{\textcolor{black}{#1}}}

\begin{document}

\title{Signature of non-trivial  band topology in Shubnikov--de Haas oscillations}

\author{Denis R. Candido}
\affiliation{Department of Physics and Astronomy, University of Iowa, Iowa City, Iowa 52242, USA}
\author{Sigurdur I. Erlingsson}
\affiliation{Department of Engineering, Reykjavik University, Menntavegi 1, IS-102 Reykjavik, Iceland}
\author{Jo\~ao Vitor I. Costa}
\affiliation{Instituto de F\'isica de S\~ao Carlos, Universidade de S\~ao Paulo, 13560-970 S\~ao Carlos, SP, Brazil}
\author{J. Carlos Egues}
\affiliation{Instituto de F\'isica de S\~ao Carlos, Universidade de S\~ao Paulo, 13560-970
S\~ao Carlos, SP, Brazil}
\affiliation{Department of Physics, University of Basel, CH-4056, Basel, Switzerland}

\begin{abstract}

\denis{We investigate the Shubnikov-de Haas (SdH) magneto-oscillations in the resistivity of two-dimensional topological insulators (TIs). 
Within the Bernevig-Hughes-Zhang (BHZ) model for TIs in the presence of a quantizing magnetic field, we 
obtain analytical expressions for the SdH oscillations by combining a semiclassical approach for the resistivity and a trace formula 
for the density of states. We show that when the non-trivial topology is produced by inverted bands with ``Mexican-hat'' shape, SdH oscillations show an anomalous beating pattern that is \textit{solely} due to the non-trivial topology of the system. These beatings are robust against, and distinct from beatings originating from spin-orbit interactions. This provides a direct way to experimentally probe the non-trivial topology of 2D TIs entirely from a bulk measurement. Furthermore, the Fourier transform of the SdH oscillations as a function of the Fermi energy and quantum capacitance models allows for extracting both the topological gap and gap at zero momentum. }

\end{abstract}

\maketitle

{\it Introduction. ---} Topological Insulators (TIs) are materials that behave as gapped insulators in bulk whereas also hosting metallic (gapless) topological helical states localized at their edges in 2D 
TIs~\cite{kane2005z2,kane2005quantum,fu2006time,bernevig2006quantumprl,bernevig2006quantum,konig2007quantum,PhysRevLett.100.236601,doi:10.1126/science.aan6003,Tang2017,candido2018} or surfaces in 3D TIs \cite{Fu2007a,Fu2007,qi2008topological,Zhang2009,liu2010model,Hughes2011}. 
For that reason, attention to topological materials has been mainly focused on edge- and surface-like phenomena. For instance, 
the corresponding experimental confirmation of TIs are usually performed via edge- or surface-related effects, e.g., the quantized 
conductivity for 2D TIs~\cite{konig2007quantum,doi:10.1126/science.1174736,10.1063/1.3577612,Knez2011}, and angle-resolved 
photoemission spectroscopy for 3D TIs~\cite{teo2008surface,Hsieh2008,chen2009experimental,Xia2009,Tang2017}. Despite 
these successful realizations, there are still open problems, e.g., the 
quantization of the resistivity is not always observed~\cite{konig2007quantum,doi:10.1126/science.1174736,PhysRevB.87.235311,PhysRevB.88.165309,PhysRevLett.114.096802,beukmann17:241401,Nichele_2016,doi:10.1126/science.aan6003,Ferreira2022}. 
This thus requires the development of alternative methods to probe the presence of 
topological bands, e.g., via the investigation of bulk properties of the material. 

\begin{figure}[t!]
\centerline{\resizebox{3.45in}{!}{\includegraphics{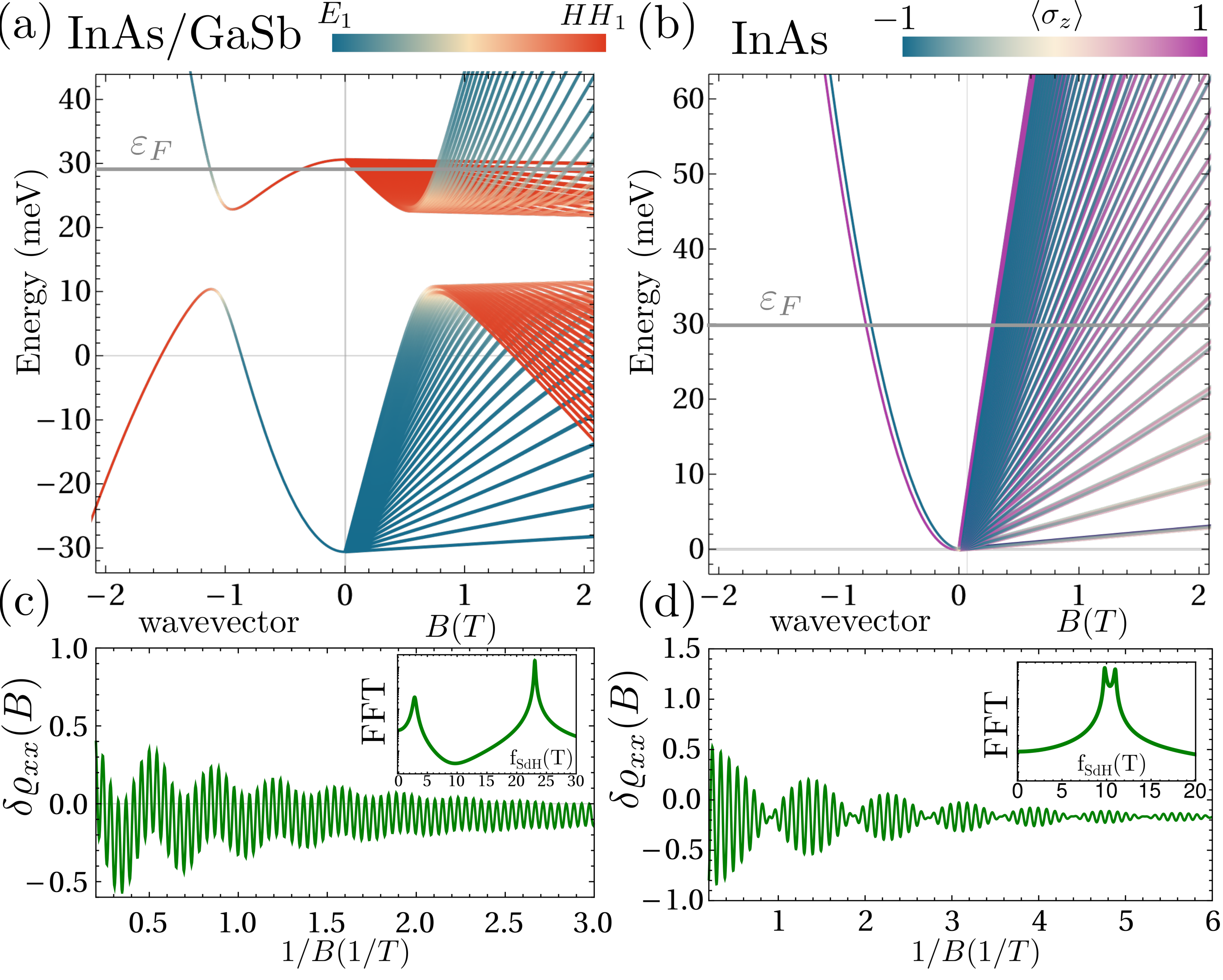}}}
\caption{(a) Energy dispersion and Laudau Levels of topological InAs/GaSb type-II quantum wells. The color code shows the $E_1$ (electron-like) and ${HH}_1$ (hole-like) band composition. (b) Same as (a) for InAs-QW with color code showing the spin expectation value. (c)  InAs/GaSb magneto-resistivity $\delta \varrho_{xx}(B)$ vs. $B$ for $\varepsilon_F=30$~meV with inset showing the corresponding Fourier transform of $\delta \varrho_{xx}(B)$. (d) Same as (c) for InAs-QW with $\varepsilon_F=30$~meV.  For all the plots we used parameters within Tab.~\ref{tab:table1}} \label{fig1}
\end{figure}

Here we first develop an analytical theory describing Shubnikov-de Haas oscillations (SdH)~\cite{sdh-original,sdh-original2,sdh-originalv2} in the magnetoresistivity of bulk 2D TIs. Using this theory, we show that topologically non-trivial bands with ``Mexican-hat'' band structure present an anomalous SdH-beating pattern not found in trivial systems, e.g., InAs quantum well (QW), Fig.~\ref{fig1}. These beatings are present for a broad Fermi energy ($\varepsilon_F$) range {in which $\varepsilon_F$ is simultaneously intersecting} both electron- and hole-like bands -- a sole characteristic of bulk insulators with band inversion. The combination of quantum capacitance models~\cite{ihn2009semiconductor,PhysRevLett.118.016801} and our SdH theory allow for the experimental determination of the Hamiltonian parameters, including both the topological gap and gap at $\mathbf{k}=0$. This is done via the extraction of the frequencies defining the corresponding SdH-oscillations as a function of the Fermi level.  We show that these anomalous beatings are fully distinguishable from the spin-orbit beatings due to Rashba and Dresselhaus spin-orbit coupling in 2D TIs. Our approach thus allows for a novel way of identifying band inversion characterizing topological insulators solely via a bulk measurement. Our method can be applied to different TIs with Mexican hat band structure, including InAs/GaSb QWs~\cite{PhysRevLett.100.236601}, strained-InSe~\cite{Ma_2013}, few layers of GaS and GaSe~\cite{PhysRevLett.108.266805}, 1T'-WTe$_2$ monolayers~\cite{Tang2017}, patterned InAs {double} QWs~\cite{PhysRevB.91.035312} and  Na$_2$XY trilayers~\cite{PhysRevB.100.205116}.

\begin{figure*}
  \includegraphics[width=\textwidth]{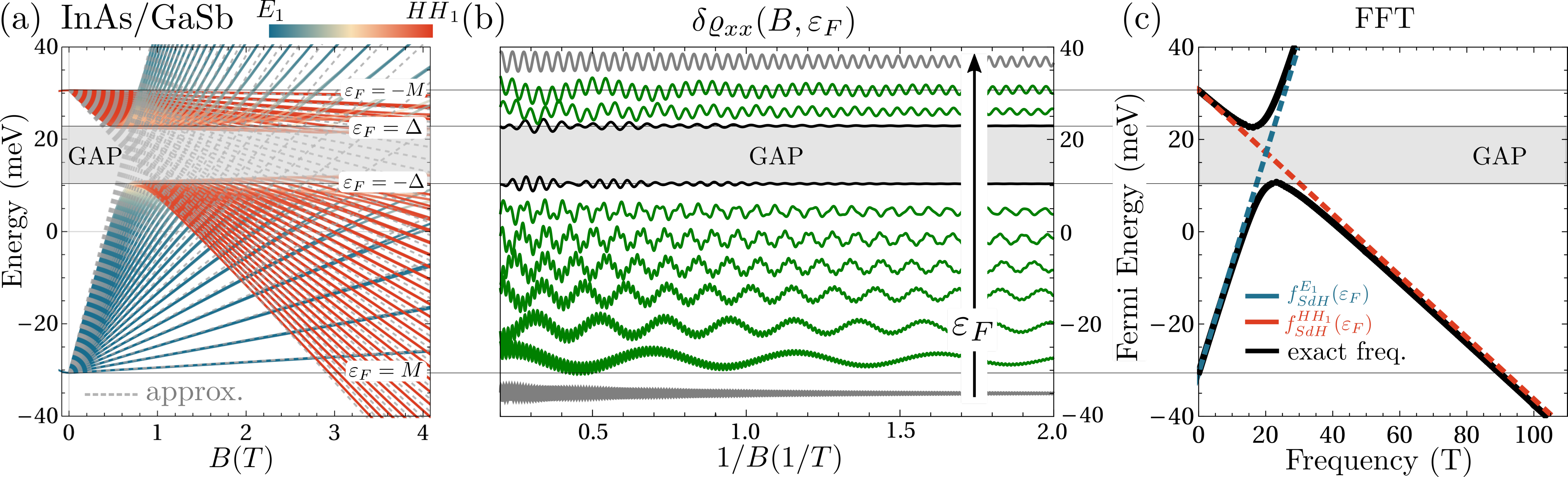}
  \caption{(a) Landau Levels [Eq.~(\ref{LL})] for InAs/GaSb QW as a function of the magnetic field $B$. Dashed gray lines are LL energies for decoupled $E_1$ and $HH_1$-like gases, i.e., Eq.~(\ref{LL}) with $A=0$. The plot is obtained with parameters from Tab.~\ref{tab:table1}. (b) Normalized magneto-oscillations in the longitudinal resistivity as a function of $1/B$ for different Fermi energies. Unusual beatings are represented by the green solid lines and are present for $-|M|\lesssim \varepsilon_F \lesssim -|\Delta|$ and $|\Delta|\lesssim \varepsilon_F \lesssim |M|$. For $\varepsilon_F\approx \pm \Delta$ the frequencies become closer and the ordinary beating pattern is found (black solid lines). For $|\varepsilon_F| \gtrsim \pm |M|$ we effectively have only one gas (either electron or hole), and we only obtain one frequency (gray solid lines). (c) Fermi energy dependence of the SdH frequencies. Black curves represent the exact SdH-frequencies arising from the FFT of Eq.~(\ref{normdiffmag}) and blue (red) dashed line represents the frequencies of a decoupled electron (hole) gas. Note the absence of any frequency for $|\varepsilon_F|\lesssim|\Delta|$ (``frequency gap''), the presence of a single frequency for $|\varepsilon|\gtrsim|M|$, and the presence of two frequencies for both $-|M|\lesssim \varepsilon_F \lesssim -|\Delta|$ and $|\Delta|\lesssim \varepsilon_F \lesssim |M|$ energy intervals.}  \label{fig2}
\end{figure*}

{\it Hamiltonian.---}
We use the Bernevig-Hughes-Zhang (BHZ) model~\cite{bernevig2006quantum} to obtain the energy dispersion of the lowest conduction ($E_1$) and the highest valence ($HH_1$) of 2D TIs, e.g., type-I HgTe/CdTe QW~\cite{bernevig2006quantum}, type-II InAs/GaSb QW~\cite{PhysRevLett.100.236601,franz2013topological}
\begin{equation}
{\cal H}\left(\mathbf{k}\right)=\left(\begin{array}{cc}
h\left(\mathbf{k}\right) & H_{SO}\left(\mathbf{k}\right)\\
H_{SO}^*\left(\mathbf{k}\right) & h^*\left(-\mathbf{k}\right)
\end{array}\right),
\label{HBHZ_k}
\end{equation}
with $h_{}\left(\mathbf{k}\right)=-D{\mathbf{k}^{2}}\mathbf{1}_{2\times2}+\mathbf{d}_{}\left(\mathbf{k}\right)\cdot\boldsymbol{\tau}$,
$\mathbf{d}_{}\left(\mathbf{k}\right)=\left( Ak_{x},\,-Ak_{y},\,M-\mathcal{B}{\mathbf{k}^{2}}\right)$,
$\mathbf{k}$ is the in-plane wave vector, $\boldsymbol{\tau}$
 the Pauli matrices describing the pseudospin space, $H_{SO}$ the Dresselhaus (bulk inversion asymmetry or BIA) and Rashba (structure inversion asymmetry asymetry, or SIA) spin-orbit Hamiltonians, and $A,{\cal B},D,M$ the effective QW $\mathbf{k}\cdot\mathbf{p}$ parameters. However, for $D=-\hbar^2/2m^{*}$, $A = \alpha$ and ${{\cal B}=M=H_{SO}(\mathbf{k})=0}$, this Hamiltonian also describes the energy dispersion of a two-dimensional electron gas (2DEG) with effective mass $m^*$ and spin-orbit Rashba parameter $\alpha$ ~[Supplemental Material~\cite{sm}]. The energy dispersions for InAs/GaSb and InAs QWs are plotted for the parameters of Table~\ref{tab:table1} in Fig.~\ref{fig1}(a) and (b), respectively, with the corresponding color map representing the contribution of $E_1$ and $HH_1$ subbands and spin value. In Fig.~\ref{fig1}(a), the InAs/GaSb spectrum contains a band inversion characterized by the higher energy of $HH_1$-subband with respect to $E_1$ at $k=0$. Due to the type-II InAs/GaSb QW structure, the overlap between $E_1$ and $HH_1$ envelope functions is small, resulting in a small gap opening via parameter $A$. The corresponding shape of the spectrum is often described as ``Mexican-hat'' and it is a characteristic of topologically non-trivial systems. The condition for this Mexican-hat regime in the conduction (valence) band is given by $2\left[{\cal B}+\textrm{sign}\left({\cal B}\right)D\right]M>A^{2}$ ($2\left[{\cal B}-\textrm{sign}\left({\cal B}\right)D\right]M>A^{2}$) with $M {\cal B}>0$ and $|{\cal B}|>|D|$ (see Supplemental Material~\cite{sm}).

In the presence of a perpendicular
magnetic field $\mathbf{B}=B_{0}\hat{z}$, using the Landau
gauge with vector potential $\textbf{A}=B_{0}\left(-y,0,0\right)$, the corresponding BHZ Hamiltonian [Eq.~(\ref{HBHZ_k})] is obtained via the minimum coupling $\mathbf{p}\rightarrow\bm{\Pi}$ with $\bm{\Pi}=\textbf{p}-q\textbf{A}$,
and $\mathbf{p}=1/(i\hbar)\boldsymbol{\nabla}$~\cite{candido2023quantum}. It is convenient to introduce the creation and annihilation bosonic operators, $a^\dagger={\ell_{c}}\left(\Pi_{x}+i\Pi_{y}\right)/{\sqrt{2}\hbar}$ and $a={\ell_{c}}\left(\Pi_{x}-i\Pi_{y}\right)/{\sqrt{2}\hbar}$, respectively, with $\left[a,a^{\dagger}\right]=1$, ${a\left|n\right\rangle =\sqrt{n}\left|n-1\right\rangle}$, and ${a^{\dagger}\left|n\right\rangle =\sqrt{n+1}\left|n+1\right\rangle}$. In leading order in the spin-orbit terms, the Hamiltonian reads
\begin{widetext}
\small{
\begin{equation}
{\cal H}=\left[\begin{array}{cccc}
\left(\hbar\omega_{1}+\hbar\omega_{2}\right)\left(a^{\dagger}a+\frac{1}{2}\right)+M & \hbar\eta a^{\dagger} & \hbar\alpha a & -\Delta_{BIA}\\
\hbar\eta a & \left(\hbar\omega_{2}-\hbar\omega_{1}\right)\left(a^{\dagger}a+\frac{1}{2}\right)-M & \Delta_{BIA} & 0\\
\hbar\alpha a^{\dagger} & \Delta_{BIA} & \left(\hbar\omega_{1}+\hbar\omega_{2}\right)\left(a^{\dagger}a+\frac{1}{2}\right)+M & -\eta a\\
-\Delta_{BIA} & 0 & -\eta a^{\dagger} & \left(\hbar\omega_{2}-\hbar\omega_{1}\right)\left(a^{\dagger}a+\frac{1}{2}\right)-M
\end{array}\right], \label{Hmag}
\end{equation}}
\end{widetext}
with $\hbar\omega_{1}=-2B/\ell_{c}^{2}$,
$\hbar\omega_{2}=-2D/\ell_{c}^{2},$ $\hbar\eta=A\sqrt{2}/\ell_{c}$, $\hbar\alpha=\alpha_{e}\sqrt{2}/\ell_{c}$ and $\ell_{c}=\sqrt{\hbar/|eB|}$. In the absence of spin-orbit terms (i.e., $\alpha=\Delta_{BIA}=0$), Hamiltonian Eq.~(\ref{Hmag}) has an analytical Landau level (LL) structure~\cite{PhysRevB.90.115305}
\begin{align}
\varepsilon_{n,\tau}^{\sigma=\pm} & =\frac{1}{2}\left[\pm\hbar\omega_{1}+2n\hbar\omega_{2}\right.\nonumber \\ 
 & +\left.\tau\sqrt{\left(\hbar\omega_{2}\pm2n\hbar\omega_{1}\pm2M\right)^{2}+4n\left(\hbar\eta\right)^{2}}\right],\label{LL}
\end{align}
with $\sigma=\pm$ ($\tau=\pm$) representing the spin (pseudospin) subspace~\footnote{Despite the common claim affirming that each BHZ block corresponds to different spin, we emphasize this is not the case as the $E_1$ subbands contain a mix between conduction band up and light hole bands down~\cite{candido2018,candido2018blurring}.}, $n\in\mathbb{N}_0$ and $\varepsilon_{n=0}^{\sigma=\pm}= \pm M+\left(\hbar\omega_{2}\pm\hbar\omega_{1}\right)/2$. In the right axis of Figs.~\ref{fig1}(a) and (b) we plot the LLs energies [Eq.~(\ref{LL})] as a function of  $B$  for both InAs/GaSb and InAs QWs~\footnote{For InAs, the LL energy expresion becomes $\varepsilon_{n,\sigma}=\hbar\omega_{2}n+\frac{\sigma}{2}\sqrt{\left(\hbar\omega_{2}\right)^{2}+4\left(\hbar\eta\right)^{2}n}$ with $\sigma=\pm$ representing the different spins (see Supplemental Material~\cite{sm}).}, respectively. While all the LLs for the InAs QW present a monotonic linear dependence on $B$, LLs for InAs/GaSb do not. It is interesting to note, however, that far from the gap region, both $E_1$-like LLs (blue color) and $HH_1$-like LLs (red color) present a linear monotonic dependence on $B$, with slopes with opposite signs. This happens since for $|\varepsilon|\gg|M|$ we have effectively decoupled the electron and hole bands (i.e., $\eta \propto A= 0$) with positive and negative effective masses, $m_e=-\hbar^2/(2D+2\mathcal{B})$ and $m_h=-\hbar^2/(2D-2\mathcal{B})$, respectively. For these regions, their energies read ${{\varepsilon_{n,E_1}^{\sigma=\pm}}\equiv \varepsilon_{n,\tau=-}^{\sigma=\pm} \approx M + \hbar |\omega_e|(n+1/2)}$ and ${{\varepsilon_{n,HH_1}^{\sigma=\pm}}\equiv{\varepsilon_{n,\tau=+}^{\sigma=\pm}} \approx -M - \hbar |\omega_h|(n+1/2)}$, with $\omega_e=eB/m_e=\omega_1+\omega_2$ and $\omega_h=eB/m_h=\omega_2-\omega_1$ [see dashed lines in Fig.~\ref{fig2}(a)]. Conversely, closer to the gap region, $E_1$- and $HH_1$-like LLs interact strongly with each other via the off-diagonal $\eta$ term, resulting in an anti-crossing of these LLs, similarly to the one obtained for the bulk bands at $B=0$. As we will present next, when the Fermi energy is simultaneously crossing electron- and hole-like bands, an anomalous SdH-oscillation appears.
\begin{table}[t!]
\caption{\label{tab:table1} BHZ parameters $A$ (meV.nm), ${\cal B},D$ (meV.nm$^2$) and $M,\Delta_{BIA}$ (meV) for InAs/GaSb-QW and InAs 2DEG. For both systems we have used $\Gamma=0.3$~meV.}
\begin{ruledtabular}
\begin{tabular}{cccccc}
& $A$  & ${\cal B}$ & $D$ & $M$&$\Delta_{BIA}$ \\
\colrule
\textrm{InAs/GaSb}\cite{PhysRevLett.118.016801} & $306$ & $-500$ & $-280$ & $-30.6$ & $0.2$ \\
\textrm{InAs-QW}\cite{beukmann17:241401} & $100$ & $0$ & $-95188$ & $0$ &$0$ 
\end{tabular} 
\end{ruledtabular}
\end{table}

\begin{figure*}
  \includegraphics[width=\textwidth]{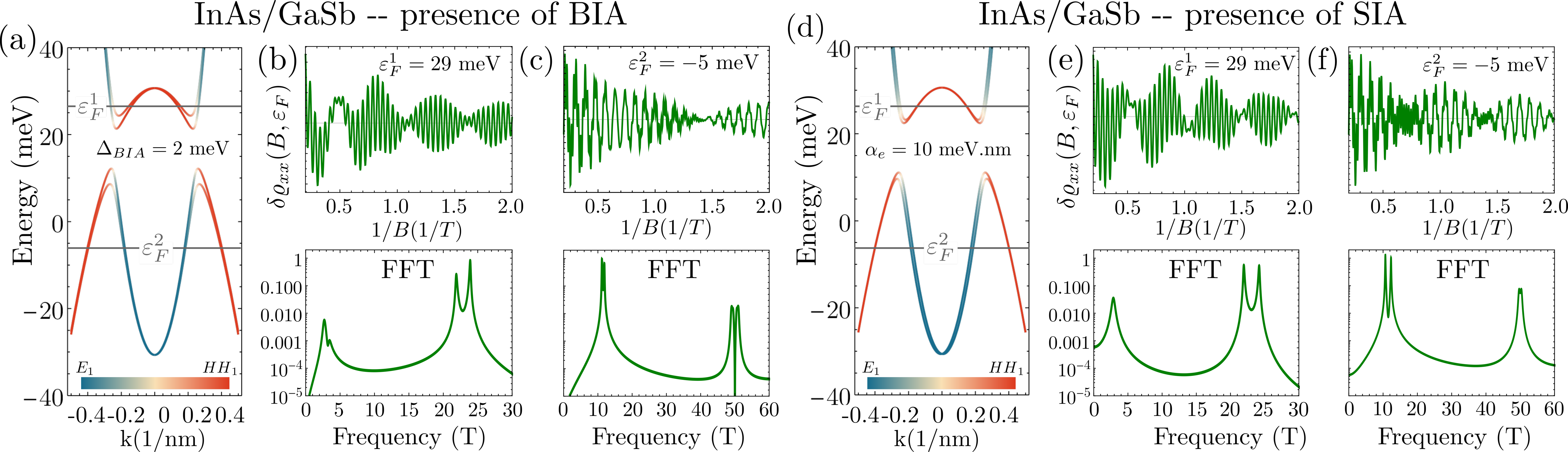}
  \caption{Effect of different spin-orbit interactions on the energy dispersions,  SdH oscillations in $\varrho_{xx}(B)$ and their corresponding Fourier spectra of InAs/GaSb QWs. (a) Energy dispersion of topological InAs/GaSb QW in the presence of BIA with $\Delta_{BIA}=2$~meV. $\delta \varrho_{xx}(B,\varepsilon_F)$ vs $1/B$ and corresponding FFT for (b) $\varepsilon_F=29$~meV and (c) $\varepsilon_F=-5$~meV. (d) Energy dispersion of topological InAs/GaSb QW in the presence of SIA with $\alpha_{e}=10$~meV.nm. $\delta\varrho_{xx}(B,\varepsilon_F)$ vs $1/B$ and corresponding FFT for (e) $\varepsilon_F=29$~meV and (f) $\varepsilon_F=-5$~meV. All the plots were generated with parameters from Tab.~\ref{tab:table1} }  \label{fig3}
\end{figure*}

{\it Shubnikov-de Haas oscillations in 2D TIs.} Magneto-oscillations in the longitudinal resistivity $\varrho_{xx}(B)$ are called SdH oscillations~\cite{sdh-original,sdh-original2,sdh-originalv2}. They arise due to the sequential crossing between the Fermi energy and the LLs of the system, which yields a corresponding depopulation of the LLs as the magnetic field is increased. The corresponding rate at which these crossings happen dictates the periodicity of the SdH oscillations with respect to $1/B$~\cite{PhysRevB.82.155456,candido2023quantum}. It is well-known that for spin-orbit coupled 2DEGs (e.g., InAs QWs), each split band [see Fig.~\ref{fig1}(b)] gives rise to SdH oscillations with similar frequencies. The sum of these oscillations, in turn, produces SdH oscillations with beatings, shown in Fig.~\ref{fig1}(d)~\cite{ihn2009semiconductor,PhysRevB.3.4274,PhysRevB.38.10142,RWollrab_1989,das89:1411,PhysRevResearch.5.043297}. 

More quantitatively, SdH oscillations in the resistivity $\varrho_{xx}$ $\varrho_{xx}(B)$
can be obtained using Drude's semi-classical equations accounting for the magnetic field dependence of the electron scattering 
time ($\tau$) via Fermi's golden rule~\cite{PhysRevResearch.5.043297}, whereas a more rigorous formalism can be found in 
Ref.~\cite{RevModPhys.84.1709}. Then, $\varrho_{xx}(B)=m^*/ne^2\tau(B)$ with $1/\tau(B)\propto \rho(\varepsilon_F,B)$, 
where $\rho(\varepsilon,B)$ is the density of states (DOS) (per spin and area) of our 2DEG, i.e., 
$\rho(\varepsilon,B)=\frac{\tilde{D}}{A}\sum_{n,\tau,\sigma}\delta(\varepsilon-\varepsilon_{n,\tau}^\sigma)$, with  $n\in\mathbb{N}_0$ 
and LL degeneracy $\tilde{D}=A/2\pi\ell_{c}^2$.
To lowest order in the deviations of the DOS $\delta\rho(\varepsilon,B)$, 
the scattering time in the presence of a finite magnetic field is $\tau(B)\approx\tau_0[1-\delta\rho(\varepsilon,B)/\rho_0]$, \
with $\delta\rho(\varepsilon,B)=\rho(\varepsilon,B)-\rho_0$~\cite{ihn2009semiconductor,candido2023quantum}, 
 $\tau_0$ and $\rho_0$ the scattering time and DOS at $B=0$, respectively. The resistivity then becomes dependent on the magnetic field 
$\varrho_{xx}(B)=\varrho_{xx}^0[1+{2}\delta\rho(\varepsilon,B)/\rho_0]$~\cite{candido2023quantum,RevModPhys.84.1709} 
where $\varrho_{xx}^0$ is the Drude's conductivity at $B=0$. To obtain the analytical formula for $\varrho_{xx}(B)$, 
we will use the formalism employed in Refs.~\cite{candido2023quantum,ihn2009semiconductor}, which makes use of the Poisson's summation formula~\cite{brack97:book}. For a Hamiltonian with corresponding pseudo-spin $\tau$ and spin index $\sigma$, the normalized magnetoresistivity $\delta\varrho_{xx}(B)=[\varrho_{xx}(B)-\varrho_{xx}^0]/\varrho_{xx}^0$ at $T=0$~K reads~\cite{candido2023quantum}
\begin{eqnarray}
\delta {\varrho}_{xx}(B)={2}\sum_{l,\sigma,\tau}^{}  \left.e^{-\pi l\Gamma |dF_{\tau}^{\sigma}/d\varepsilon|}
\cos\left[2\pi lF_{\tau}^{\sigma}(\varepsilon)\right]\right|_{\varepsilon=\varepsilon_F},\label{normdiffmag}
\end{eqnarray}
with $l\in\mathbb{N}$ representing the different harmonics, $\Gamma$ the LL broadening, and the $F$-functions defined by ${\varepsilon_{n,\tau}^{\sigma}=\varepsilon\leftrightarrow n=F_{\tau}^{\sigma}(\varepsilon)}$, with $\varepsilon_{n,\tau}^{\sigma}$ [Eq.~(\ref{LL})] determined from the BHZ Hamiltonian. Thus, to calculate the magneto-resistivity we have to first find $F_{\tau}^\sigma(\varepsilon)$ by inverting the LL expression. In the Supplemental Material~\cite{sm} we present the (lengthy) analytical formula for $F_{\tau}^\sigma(\varepsilon)$.


We start by discussing SdH oscillations for a InAs 2DEG with parameters in Table~\ref{tab:table1}. Due to the dominance of the quadratic term in $\mathbf{k}$ over the linear term, Eq.~(\ref{normdiffmag}) yields SdH-oscillations due to two cosines with similar F-fuctions $F_{+}^{\sigma}(\varepsilon_F)$ and $F_{-}^{\sigma}(\varepsilon_F)$ (Supplemental Material~\cite{sm}). Consequently, the total resistivity is a sum of oscillations with similar frequencies,
thus presenting a beating pattern, shown in Fig.~\ref{fig1}(d). 
The faster frequency is proportional to the 2DEG effective mass $m^*$ and $\varepsilon_F$ via, $F_{+}^{\sigma}(\varepsilon_F)+F_{-}^{\sigma}(\varepsilon_F) = \frac{\varepsilon_F m^*}{e\hbar} \frac{1}{B}$, while the slower one (defining the beatings) is $F_{+}^{\sigma}(\varepsilon_F)-F_{-}^{\sigma}(\varepsilon_F)\propto \alpha/B$~\cite{candido2023quantum}, with Rashba spin-orbit parameter $\alpha$. The Fourier transform of the corresponding SdH-oscillation is plotted in the inset of Fig.~\ref{fig1}(d), where we observe similar SdH frequencies around $\sim 10$~T. Surprisingly,  InAs/GaSb QW in the topological regime presents SdH oscillations with a completely different pattern, shown in Fig.~\ref{fig1}(c). Its Fourier transform shows that instead of similar frequencies yielding a beating pattern, this system contains two frequencies with very different values, $\sim 3$~T and $\sim 23$~T [see inset Fig.~\ref{fig1}(c)]. This happens due to the current being carried by electron- and hole-like bands with contrasting Fermi areas. As we explain below, this arises solely due to the non-trivial topology of the system emerging from the band inversion between $E_1$ and $HH_1$ subbands.

In Fig.~\ref{fig2}(b) we plot $\delta\varrho_{xx}(B)$ versus $1/B$ for different Fermi energy values (right $y$-axis). The anomalous beatings (green curves) are present for a wide range of $\varepsilon_F$ where $E_1$- and $HH_1$-like bands coexist with the same energy, i.e., $|\Delta| \lesssim |\varepsilon_F| \lesssim |M|$. As already discussed, far from the anti-crossing region between $E_1$ and $HH_1$ ($|\varepsilon_F|\gg|M|$), the eigenenergies [Eq.~(\ref{LL})] can be described by LLs of decoupled $E_1$-electron and $HH_1$-hole gases, with F-functions $F_{E_1} (\varepsilon_F) \equiv F_{\tau=-}^{\sigma}{(\varepsilon_F)}\approx (\varepsilon_F-M)/\hbar\omega_e -1/2$ and $F_{HH_1}{(\varepsilon_F)}\equiv F_{\tau=+}^{\sigma} (\varepsilon_F)\approx - (\varepsilon_F+M)/\hbar\omega_h +1/2 $, respectively, and corresponding SdH frequencies  $f_{SdH}^{E_1}= \varepsilon_F m_e/\hbar e$ and $f_{SdH}^{HH_1}=-\varepsilon_F |m_h|/\hbar e$, plotted as blue and red dashed lines in Fig.~\ref{fig2}(c). In contrast to the InAs case, these frequencies are now completely distinct from each other, which is a consequence of the striking difference in the electron and hole effective masses of InAs/GaSb QW. Furthermore, due to the different signs of the $m_e$ and $m_h$ effective masses, these frequencies possess opposite dependence on $\varepsilon_F$, i.e., while $f_{SdH}^{E_1}$ increases with $\varepsilon_F$, $f_{SdH}^{HH_1}$ decreases. As a consequence, increasing $|\varepsilon_F|$ yields an increase in the frequency separation [see Fig.~\ref{fig2}(c)].  Interestingly, $f_{SdH}^{E_1}\rightarrow0$ ($f_{SdH}^{HH_1}\rightarrow0$) for $\varepsilon_F \rightarrow M$ ($\varepsilon_F \rightarrow -M$), which is follows from the absence of $HH_1$-like ($E_1$-like) states for $\varepsilon_F>-M$ ($\varepsilon_F< M$). Therefore, we only have one frequency for $|\varepsilon_F| > |M|$, and beating is absent [see gray {area} in Fig.~\ref{fig2}(b)]. Around $\varepsilon_F\approx \pm \Delta$ the frequencies become comparable and we observe the usual beating pattern of a 2DEG with spin-orbit coupling [see black curve in Fig.~\ref{fig2}(b)]. Finally, the F-functions do not exist within the gap $-|\Delta|\lesssim \varepsilon_F \lesssim |\Delta|$. 

{\it Experimental realization. ---} 
Here we demonstrate how the experimental study of the SdH oscillations versus $\varepsilon_F$ in TIs with Mexican-hat shaped bands allows for the reconstruction of their bulk bands, and corresponding confirmation of its non-trivial topology. A top gate $V_g$ applied to the system can control the Fermi level, with the quantum capacitance formula permitting us to translate voltage values to the corresponding Fermi level, i.e., $\varepsilon_F=\varepsilon_F(V_g)$~\cite{ihn2009semiconductor,PhysRevLett.118.016801}. First, by increasing the voltage such that $\varepsilon_F\gg|M|$, we will obtain magneto oscillations solely due to $E_1$-states with corresponding frequency $f_{SdH}^{E_1}$. Performing the FFT of these oscillations permits us to extract the effective mass of the $E_1$-states as $f_{SdH}^{E_1}=\varepsilon_F m_e/\hbar e$. The same analysis can be done for negative $V_g$ such that $\varepsilon_F\ll-|M|$, and the effective mass of $HH_1$-states can also obtained. As we diminish $\varepsilon_F$ via $V_g$, the oscillations display and additional frequency once $\varepsilon_F\lesssim-M$ is reached, thus allowing us to obtain the parameter $M$ defining the gap at $k=0$. Both frequencies will vanish for $\varepsilon_F\rightarrow|\Delta|$, and we can extract the value of the bulk topological gap $2\Delta$. With all these parameters we are able to fully reconstruct the TI band structure and its corresponding topology. A few works have already performed measurement in similar systems~\cite{PhysRevLett.122.186802,PhysRevB.99.201402,PhysRevLett.118.016801,PhysRevB.104.085301}. However, here we provide a systematic way of probing the non-trivial band topology and extracting the corresponding effective parameters of the 2D TI.


{\it Robustness against spin-orbit coupling. ---} 
To further test the robustness of the anomolous beatings as evidence of the non-trivial topology of the system, we study SdH oscillations in the presence of different spin-orbit couplings $\Delta_{BIA}$ and $\alpha$ [See Eq.~(\ref{Hmag})]. The presence of $\Delta_{BIA}$ term breaks the spin degeneracy of both conduction and valance bands, plotted in Fig.~\ref{fig3}(a) for $\Delta_{BIA}=2$~meV. The corresponding SdH oscillations and its Fourier transform are plotted in Fig.~\ref{fig3}(b) and (c) for $\varepsilon_F=29$~meV and $\varepsilon_F=-5$~meV, respectively. Compared to the case with $\Delta_{BIA}=0$, shown in Fig.~\ref{fig1}(c), here we obtain a slightly different beating pattern. Nevertheless, the corresponding FFT plot shows that $\Delta_{BIA}$ does not shift the main frequency peaks centered at $\sim3$~T and $\sim23$~T, but rather split them, similarly to the case of Rashba SO in 2DEG, shown in Fig.~\ref{fig1}(d). In Fig.~\ref{fig3}(d) we present results for the SIA term $\alpha_e=10$~meV.nm, which produces similar spin split features. However, for $\varepsilon_F=29$~meV ($
\varepsilon{_F}=-5$~meV) this term splits mainly the higher (lower) SdH frequency peak, as $\alpha_e$ couples different spin components of the $E_1$-subbands, i.e., the band responsible for oscillations with higher (low) frequency for $\varepsilon_F=29$~meV ($\varepsilon{_F}=-5$~meV). In short, the results summarized in Fig.~\ref{fig3} show that the presence of different spin-orbit coupling does not alter nor does it change the main features of the anomalous SdH-oscillations and its frequencies, thus proving that the non-trivial band topology of the system can be inferred from SdH bulk transport measurements.

{\it Conclusion. ---} 
We have developed an analytical formalism to describe the SdH-oscillations of 2D TIs with Mexican hat band structure. Their SdH-oscillations contain anomalous beatings that are very distinct from the ones found in ordinary trivial semiconductors in the presence of spin-orbit coupling. These beatings are originated from two contrasting frequencies that arise from the presence of overlaping electron and hole-like Fermi surfaces, a unique characteristic of inverted band insulators with Mexican hat band structure. Finally, we show that an analysis of the frequencies versus the Fermi energy allows for a straightforward extraction of the BHZ Hamiltonian parameters, including the topological gap. Our work thus demonstrates an alternative method -- based solely on bulk properties -- for demonstrating the manifestation of the non-trivial topology of 2D TIs.


 {\it Acknowledgments.---} DRC acknowledges funding from the University of Iowa Fund. SIE was supported by the Reykjavik University Research Fund. JCE acknowledges funding from the National Council for Scientific and Technological Development (CNPq) Grant No. 301595/2022-4 and São Paulo Research Foundation (FAPESP), Grant 2020/00841-9.



\begin{acknowledgments}
\end{acknowledgments}

\bibliography{references}

\begin{thebibliography}{56}
\bibitem [{sm()}]{sm}%
  \BibitemOpen
  \href@noop {} {\bibinfo  {journal} {See {S}upplemental {M}aterial at [URL
  will be inserted by publisher] for the derivation of the {L}indblad
  equation}\ }\BibitemShut {NoStop}%
\end{thebibliography}

\end{document}